\documentclass[12pt,amsmath,amssymb]{revtex4}

\begin{document}

\title{Singularity formation on a fluid interface during the Kelvin-Helmholtz instability development }

\author{ N.M. Zubarev$^{(a,b)}$, E.A. Kuznetsov$^{(b,c,d)}$\/\thanks{kuznetso@itp.ac.ru}}
\affiliation{{\small $^{(a)}$ Institute of Electrophysics of the Ural Branch of the Russian Academy of Sciences\ 
620016 Ekaterinburg, Russia}\\
{\small $^{(b)}$ P.N. Lebedev Physical Institute of the Russian Academy of Sciences\ 
119991 Moscow, Russia} \\
{\small $^{(c)}$ Novosibirsk State University\ 
630090 Novosibirsk, Russia}\\
{\small $^{(d)}$ L.D. Landau Institute for Theoretical Physics of the Russian Academy of Sciences\ 
142432, Chernogolovka, Moscow region, Russia}}

\begin{abstract}
The dynamics of singularity formation on the interface between two ideal fluids is studied for the Kelvin-Helmholtz instability development within the Hamiltonian formalism. It is shown that the equations of motion derived  in the small interface angle approximation  (gravity and capillary forces are neglected) admit exact solutions in the implicit form.
The analysis of these solutions shows that, in the general case, weak root singularities are formed on the interface in a finite time for which the curvature becomes infinite, 
while the slope angles remain small. 
For Atwood numbers close to unity in absolute values, the surface curvature has a definite sign correlated with the boundary deformation directed towards the light fluid. 
For fluids with comparable densities, the curvature changes its sign in a singular point. In the particular case of fluids with equal densities, 
the obtained results are consistent with those obtained by Moore based on the Birkhoff-Rott equation analysis.  
\end{abstract}

\maketitle

\section{Introduction}

As is known, the interface between two fluids is unstable in the presence of a tangential velocity discontinuity. 
This instability, called the Kelvin-Helmholtz instability, is one of the most important in fluid dynamics. In the absence of gravity and capillary forces, the growth rate of this instability increases linearly with increasing wavenumber $k$. 
In the presence of gravity and capillary forces, which tend to return the surface to the plane state, the growth rate is strongly reconstructed \cite{lamb,lan-lif,step-fab} that, in particular, leads to the appearance of a threshold. 
For waves exited by wind, the threshold velocity value  for the Kelvin-Helmholz instability is defined by the condition of minimality of the phase velocity for the gravity-capillary waves (in the wind absence). 
The threshold value of the wind velocity is of the order of 6~m/s. 
For this wind velocity, the mass appearance of wave crests is observed due to the wave breaking process. 
The explanation for this effect relates to the wave collapse phenomenon. 
At small excesses above the instability threshold, a narrow wave packet is excited in the $k$-space. 
As is shown in Ref.~\cite{kuz-lush}, the envelope of this packet satisfies the relativistic Klein-Gordon equation with the negative square of the mass and the ``irregular'' nonlinearity, which does not stabilize the instability and, on the contrary, leads to the explosive process of amplitude growth --- the wave collapse. 

The nonlinear behavior of the interface is also rather nontrivial for this instability when the gravity and capillary forces are neglected, i.e., if the motion of fluids is governed only by the hydrodynamic pressure. 
The first important results in this field were presented in 1979 by Moore \cite{moore}. 
In this work, for the case of one fluid, it was established that the nonlinear stage of the Kelvin-Helmholtz instability is accompanied by developing singularities on the surface of the tangential discontinuity or, as saying, in the presence of a vortex sheet (the tangential velocity discontinuity corresponds to the flat infinitely thin layer with constant vorticity). 
In this case the vortex sheet motion  can be described in the frame of the integral Birkhoff-Rott equation (see, e.g., Ref.~\cite{saf}). 
The analysis of this equation \cite{moore} showed that the singularity appearance  on the discontinuity surface occurs in a finite time. 
Weak singularities arise on the discontinuity surface: for them, the surface itself remains smooth, but its curvature becomes singular.  
Later on, numerous  computer simulations \cite{baker1,kras2,shel} confirmed this analytical result of Moore. 
It was established that these singularities become  seeds for the vortex spirals centers  (see, for example, Refs.~\cite{kras1,verga}). 

In the case of an interface --- the boundary between two different fluids (i.e., for arbitrary Atwood numbers), it was shown in Refs.~\cite{baker2,baker3}, that the boundary motion  can be also described by the Birkhoff-Rott equation together with one additional equation which determines the vorticity evolution. 
Such an approach with different variations was used for numerical simulation of the Rayleigh-Taylor instability and the Richtmyer-Meshkov intability as well (see the works \cite{kerr,rm} and the references therein). 
As for analytical studies, the work \cite{bak} should be mentioned where the attempt was made to generalize the Moore's results to the case of fluids with different densities. 
There the situation was considered where a tangential velocity discontinuity emerged as a result of the  Rayleigh-Taylor instability development, namely, when the Kelvin-Helmholtz instability was secondary. 
In this case, a tendency to the formation of singularities of the Moore's type was also demonstrated, however, in the analysis of the motion equations there was used a number of approximations which, in our opinion,  require an additional investigation. 
The most essential assumption is the so-called ``localized approximation'', which permits the authors of \cite{bak,caf}) to find an exact solution to the simplified equations of motion. 
The idea of these simplifications was to neglect the nonlinear cross-terms in the equations written in terms of analytic continuations into the upper and lower half-planes of the corresponding complex variable. 
This procedure allowed to reduce the equations to the local form, which does not contain nonlocal integro-differential operators. 
However, the neglected nonlinear interaction, generally speaking, is not small compared to the local one. 
In the present work, we will study this point in more detail; it will be shown that, for the proper choice of variables, cross-terms disappear in a natural way.  

Unlike Refs.~\cite{bak,caf}, our description is based on the Hamiltonian formalism, which was successfully applied to the similar problems concerning  the formation of singularities on the free surface \cite{kuz2} and on the interface of two ideal fluids as well \cite{kuz1}. 
It will be demonstrated that the singularities emerging due to the Kelvin-Helmholz instability has a root behavior: the surface profile  and its first derivative occur continuous functions, but the second derivative becomes infinite while approaching the collapse instant. 
In Refs.~\cite{kuz2,kuz1}, such singularities were called ``weak''. 
Their formation can be analytically described with the help of the small-angle approximation and an appropriate analytic continuation. 
In the present work it is shown that, for the Atwood numbers close to unity in absolute value, the curvature of the surface near the singularity has a definite sign (there is a tendency for the deformation of the boundary towards the light fluid). 
For fluids with comparable densities, the curvature changes its sign at the singular point. 
In the particular case of fluids with equal densities, the obtained results are consistent with those obtained by Moore on the base of the Birkhoff-Rott equation analysis.

\section{Original equations}

Let us consider two ideal incompressible fluids. Their flows will be assumed  to be potential. We restrict our consideration to two-dimensional flows when all quantities depend only on two spatial coordinates $x$ and $y$. The $y$ axis is directed normally to the unperturbed (flat) interface. Let the interface be given by  function $y=\eta (x,t)$ ($\eta =0$ in the unperturbed state).

Because of incompressibility, the velocity potentials for both fluids, $\Phi _{1,2}$, satisfy Laplace equations (here and below the subscripts \textquotedblleft 1\textquotedblright\ and \textquotedblleft 2\textquotedblright\ refer to the lower and upper fluids, respectively): 
\[
\nabla ^{2}\Phi _{1}=0,\qquad y<\eta (x,t), 
\]
\[
\nabla ^{2}\Phi _{2}=0,\qquad y>\eta (x,t). 
\]

At an infinite distance from the boundary, $y\rightarrow \mp \infty,$ the velocity fields are supposed to be uniform: $\Phi _{1,2}\rightarrow V_{1,2}x,$ where $V_{1,2}$ are corresponding constant velocities at infinity.

For the further analysis it is convenient to consider fluid  motions  in the center-of-mass frame, i.e., when the following condition is satisfied, 
\begin{equation}
\rho _{1}V_{1}+\rho _{2}V_{2}=0,  \label{mass}
\end{equation}
where $\rho _{1}$ and $\rho _{2}$ are the densities of fluids. For certainty, we will assume that $V_{1}$ is positive. Then, according to 
(\ref{mass}), $V_{2}=-(\rho _{1}/\rho _{2})V_{1}<0$.

The interface  motion is determined by the dynamic and kinematic boundary conditions:
\begin{equation}
\rho _{1}\left( \frac{\partial \Phi _{1}}{\partial t}+\frac{(\nabla \Phi
_{1})^{2}}{2}\right) -\rho _{1}\left( \frac{\partial \Phi _{2}}{\partial 
t}+\frac{(\nabla \Phi _{2})^{2}}{2}\right) =\frac{\rho _{1}V_{1}^{2}-\rho
_{2}V_{2}^{2}}{2},\qquad \quad y=\eta (x,t),  \label{eqdyn}
\end{equation}
\begin{equation}
\eta _{t}=\partial _{n}\Phi _{1}\,\sqrt{1+\eta _{x}^{2}}=\partial _{n}\Phi
_{2}\,\sqrt{1+\eta _{x}^{2}},\quad \quad y=\eta (x,t),  \label{eqkin}
\end{equation}
where $\partial _{n}$ denotes the derivative along the boundary normal: 
\[
\partial _{n}=\frac{\partial _{y}-\eta _{x}\cdot \partial _{x}}{\sqrt{1+\eta
_{x}^{2}}}. 
\]
The equation \eqref{eqdyn} is a consequence of the pressures equality
on the interface and of the nonstationary Bernoulli equations. It does not contain the terms responsible for the gravity and capillary forces. The equation \eqref{eqkin} shows that, at $y=\eta (x,t)$, the normal components of velocities for both fluids coincide,  represented the surface velocity.

It is convenient to introduce the auxiliary velocity potentials, 
\[
\tilde{\Phi}_{1,2}=\Phi _{1,2}-V_{1,2}x, 
\]
which are identically equal to zero in the unperturbed state. Let  functions  $\psi _{1,2}$ be the limiting values of the corresponding velocity potentials, $\left. \tilde{\Phi}_{1,2}\right\vert _{S}$, at the interface. 
We also introduce a new variable 
$\psi(x,t)\equiv \rho _{1}\psi _{1}-\rho _{2}\psi _{2}$. As is shown in Ref.~\cite{HAM}, the equations for the surface motion take the Hamiltonian form, 
\[
\psi _{t}=-\frac{\delta H}{\delta \eta },\quad \quad \eta _{t}=\frac{\delta 
H}{\delta \psi }, \]
where the Hamiltonian coincides with the total energy of the system,
\begin{equation}
H=\rho _{1}\iint\nolimits_{y\leq \eta }\frac{(\nabla \Phi 
_{1})^{2}-V_{1}^{2}}{2}dx\,dy+\rho _{2}\iint\nolimits_{y\geq \eta }\frac{(\nabla \Phi
_{2})^{2}-V_{2}^{2}}{2}dx\,dy.  \label{eqham1}
\end{equation}
These variables, $\psi (x,t)$ and $\eta (x,t)$, generalize the canonical variables introduced by V.E.~Zakharov for the surface waves 
\cite{zah} (see also Ref.~\cite{ZakharovKuznetsov}).

\section{Weakly nonlinear equations of motion}

Consider behavior of the system in the approximation of small angles of inclination of the boundary, when $|\eta _{x}|\ll 1$. For this case it is convenient to represent  Hamiltonian \eqref{eqham1} as an integral over the interface,
\begin{equation}
H=\int_{S}\left[ \frac{1}{2}\psi \partial _{n}\Phi_{1}-\frac{\rho
_{1}V_{1}\left( \psi _{1}+\psi _{2}\right) \eta _{x}}{2\sqrt{1+\eta 
_{x}^{2}}}\right] \,dS,  \label{eqham2}
\end{equation}
where $dS$ is the surface differential, and then to expand $H$ in a power series relative to canonical variables $\psi$ and $\eta$, 
\begin{equation}
H=H_{0}+H_{int}.  \label{eqham3}
\end{equation}
Here $H_{0}$ is the Hamiltonian quadratic with respect to $\psi $ and $\eta$, $H_{int}$ is the interaction Hamiltonian, which expansion starts from  cubic term $H_{3}$. In order to find the first two terms of $H$, one should expand    quantities $\partial _{n}\Phi_{1}$, $\psi_{1}$, and $\psi _{2}$ in the Hamiltonian \eqref{eqham2} 
up to quadratic terms with respect to $\psi $ and $\eta $. Taking into account that $\tilde{\Phi}_{1,2}$ are harmonic functions, vanishing at $y\rightarrow \mp \infty,$ and, according to \eqref{eqkin}, satisfying the condition 
\[
\partial _{n}\tilde{\Phi}_{1}-\partial _{n}\tilde{\Phi}_{2}=\frac{V_{1}(\rho
_{1}+\rho _{2})\,\eta _{x}}{\rho _{2}\sqrt{1+\eta _{x}^{2}}},\quad \quad
y=\eta (x,t), 
\]
it is rather easy to find the corresponding expansions:
\[
\partial _{n}\Phi_{1}\approx \hat{k}\psi _{1}-V_1 \eta_x-(\eta \psi 
_{1x})_{x}-\hat{k}(\eta \hat{k}\psi _{1}), \]\begin{equation}
\psi _{1}\approx \frac{\psi }{\rho _{1}+\rho _{2}}+V_{1}\hat{H}\eta 
+\frac{2\rho _{2}}{(\rho _{1}+\rho _{2})^{2}}\left[ \eta \hat{k}\psi +\hat{H}(\eta
\psi _{x})\right] -V_{1}A\left[ \eta \eta _{x}-\hat{H}(\eta \hat{k}\eta 
)\right] ,  \label{exp-psi-1}
\end{equation}\[
\psi _{2}\approx -\frac{\psi }{\rho _{1}+\rho _{2}}+\frac{\rho _{1}}{\rho
_{2}}V_{1}\hat{H}\eta +\frac{2\rho _{1}}{(\rho _{1}+\rho _{2})^{2}}\left[
\eta \hat{k}\psi +\hat{H}(\eta \psi _{x})\right] -\frac{\rho _{1}}{\rho 
_{2}}V_{1}A\left[ \eta \eta _{x}-\hat{H}(\eta \hat{k}\eta )\right] . \]
Here $\hat{k}\equiv -\partial _{x}\hat{H}$ is the integral operator with the Fourier transform equal to $|k|$; $\hat{H}$ is the Hilbert transform: 
\[
\hat{H}F(x)=\frac{1}{\pi }\mathrm{p.v.}\int_{-\infty }^{+\infty 
}\frac{F(x^{\prime })}{x-x^{\prime }}\,dx^{\prime }, \]
and $A=(\rho _{1}-\rho _{2})/(\rho _{1}+\rho _{2})$  the Atwood number. 
By restricting  linear terms in the expressions (\ref{exp-psi-1}), after the substitution into the Hamiltonian one can get: 
\[
H_{0}=\frac{1}{2(\rho _{1}+\rho _{2})}\int \psi \hat{k}\psi 
\,dx-\frac{V^{2}\rho _{1}(\rho _{1}+\rho _{2})}{2\rho _{2}}\int \eta \hat{k}\eta \,dx, 
\]
whence  the dispersion relation for the Kelvin-Helmholz instability follows immediately: $\omega ^{2}=-c^{2}k^{2}<0$, where $\omega $ is the frequency, $k$ is the wave number, and $c=V_{1}\sqrt{\rho _{1}/\rho _{2}}=- V_{2}\sqrt{\rho _{2}/\rho _{1}}$. This instability is aperiodic. Recall that we use the center-of-mass frame: if the quantities $V_{1,2}$ do not satisfy  condition (\ref{mass}),  the eigen-frequency would have a real part (see, for example, Ref.~\cite{lan-lif}). Note also, that in the limit of $\rho _{1}\rightarrow 0$ (or $\rho _{2}\rightarrow 0$), the parameter $c$ will be finite if 
$V_{1}\rightarrow \infty $ as $\rho _{1}^{-1/2}$ (or, respectively, $V_{2}\rightarrow -\infty $ as $-\rho _{2}^{-1/2}$). Below we consider the limiting cases where the Atwood number $A$ takes the values $\pm 1$ when $c$ remains finite.

We switch to dimensionless variables in Eqs.~\eqref{eqham2} and \eqref{eqham3},  
\[
\psi \rightarrow \psi \cdot c\,\lambda (\rho _{1}+\rho _{2}),\quad \quad
\eta \rightarrow \eta \cdot \,\lambda ,\quad \quad t\rightarrow {t\cdot
\,\lambda \mathord{\left/
{\vphantom {t\cdot \, \lambda  c}} \right. 
\kern-\nulldelimiterspace}c},\quad \quad x\rightarrow x\cdot \,\lambda , 
\]
where $\lambda $ is the characteristic spatial scale. As a result, the Hamiltonian $H=H_{0}+H_{3}$  can be rewritten as
\[
H=\frac{1}{2}\int \left[ \psi \hat{k}\psi -\eta \hat{k}\eta \right] 
dx+\frac{A}{2}\int \eta \left[ (\psi _{x})^{2}-(\hat{k}\psi )^{2}+(\eta 
_{x})^{2}-(\hat{k}\eta )^{2}\right] dx \]
\[
-\sqrt{1-A^{2}}\int \eta \left[ \eta _{x}\hat{k}\psi +\psi _{x}\hat{k}\eta 
\right] dx. \]
and the corresponding equations of motion take the form 
\begin{eqnarray}
\psi _{t}-\hat{k}\eta &=&-\sqrt{1-A^{2}}\left[ \eta \hat{k}\psi _{x}-\psi
_{x}\hat{k}\eta -\hat{k}(\eta \psi _{x})\right]  \label{eqbase1} \\
&&+\frac{A}{2}\left[ (\hat{k}\psi )^{2}-(\psi _{x})^{2}+(\hat{k}\eta
)^{2}-(\eta _{x})^{2}+2(\eta \eta _{x})_{x}+2\hat{k}(\eta \hat{k}\eta 
)\right] ,  \nonumber
\end{eqnarray}
\begin{equation}
\eta _{t}-\hat{k}\psi =-\sqrt{1-A^{2}}\left[ \hat{k}(\eta \eta _{x})-(\eta 
\hat{k}\eta )_{x}\right] -A\left[ (\eta \psi _{x})_{x}+\hat{k}(\eta 
\hat{k}\psi )\right] .  \label{eqbase2}
\end{equation}

\section{Reduced equations of motion}

Let us perform a canonical transformation from the variables $\psi$ and $\eta$ to new ones 
\[
f=(\psi +\eta )/2,\qquad g=(\psi -\eta )/2. \] 
The equations of motion in terms of variables $f$ and $g$ preserves the Hamiltonian form  
\[
f_{t}=\frac{\delta H}{\delta g},\qquad g_{t}=-\frac{\delta H}{\delta f}, 
\] 
and the Hamiltonian transforms as $H\rightarrow 2H$. In terms of the new variables, $H$ can be written as follows:
\begin{eqnarray}
H &=&\int f\hat{k}g\,dx+(A/2)\int (f-g)\left[ 
(f_{x})^{2}-(\hat{k}f)^{2}+(g_{x})^{2}-(\hat{k}g)^{2}\right] dx  \label{H-red} \\
&&-\sqrt{1-A^{2}}\int (f-g)\left[ f_{x}\hat{k}f-g_{x}\hat{k}g\right] dx. 
\nonumber
\end{eqnarray}
In this case the linearized equations of motion are transformed into  two separated equations, 
\[
f_{t}-\hat{k}f=0,\qquad g_{t}+\hat{k}g=0. 
\]
The equation for $f$ describes exponential growth of perturbations, while the equation for $g$ describes their damping. Hence, at times of  order of the inverse growth rate,  function $g$ can be considered  small in comparison with $f$, and the quadratic and cubic terms with respect to $g$ can be neglected in the Hamiltonian (\ref{H-red}). Then 
\begin{eqnarray*}
H &=&\int f\hat{k}g\,dx+(A/2)\int (f-g)\left[ 
(f_{x})^{2}-(\hat{k}f)^{2}\right] dx \\
&&-\sqrt{1-A^{2}}\int (f-g)\left[ f_{x}\hat{k}f\right] dx,
\end{eqnarray*}
and the corresponding equations of motion have the form 
\begin{equation}
f_{t}-\hat{k}f=(A/2)\left[ (\hat{k}f)^{2}-(f_{x})^{2}\right] 
+\sqrt{1-A^{2}}\left[ f_{x}\hat{k}f\right] ,  \label{eqkey1}
\end{equation}\begin{eqnarray}
g_{t}+\hat{k}g &=&(A/2)\left[ 
(\hat{k}f)^{2}-(f_{x})^{2}+2(ff_{x})_{x}+2\hat{k}(f\hat{k}f)\right] +\sqrt{1-A^{2}}\left[ 
\hat{k}(ff_{x})-f\hat{k}f_{x}\right]  \label{eqbase22} \\
&&-A\left[ (gf_{x})_{x}+\hat{k}(g\hat{k}f)\right] -\sqrt{1-A^{2}}\left[ 
\hat{k}(gf_{x})-(g\hat{k}f)_{x}\right] .  \nonumber
\end{eqnarray}
The obtained system has one important feature: Eq. \eqref{eqkey1} for $f$ is autonomous, while the equation \eqref{eqbase22} is linear with respect to $g$. As will be shown in Section~5, Eq.~\eqref{eqkey1} can be integrated: its solution can be found in an implicit form. As for the equation \eqref{eqbase22}, it admits further simplification. Since $g$ is considered to be small as compared with $f$, all terms containing $g$ in the right-hand side of Eq.~\eqref{eqbase22} should be neglected, 
\begin{equation}
g_{t}+\hat{k}g=(A/2)\left[ 
(\hat{k}f)^{2}-(f_{x})^{2}+2(ff_{x})_{x}+2\hat{k}(f\hat{k}f)\right] +\sqrt{1-A^{2}}\left[ \hat{k}(ff_{x})-f\hat{k}f_{x}\right].  
\label{eqrel1}
\end{equation}
The solution to this equation (it is linear with respect to $g$) can be represented as a sum of a general solution $g_{0}$ of the homogeneous linear equation and a particular solution of the inhomogeneous one. The solution $g_{0}$, obviously, decays, and the particular solution corresponds to the contribution induced by $f$, due to the quadratic nonlinearity. Then the growth of $f$ results in increase of $g$.  The induced part has the order of magnitude $g=O(f^{2})$ (for details see  Sec.~5).

Thus, the original system of equations \eqref{eqbase1} and 
\eqref{eqbase2}, possessing two branches of solutions both increasing and decreasing with time can be reduced to the much more simple equations \eqref{eqkey1} and \eqref{eqrel1}, one of which is autonomous and the other describes dynamics of $g$ completely determined by the increasing mode behavior.

\section{Solution of equations of motion}

Now we consider the key equation \eqref{eqkey1} corresponding to  increasing in time solutions. We will seek for  solution of the equation as an expansion
\[
f=f_{+}+f_{-}, 
\]
where $f_{\pm }$ are the analytic continuations of the function $f$ into the upper and, respectively, lower half-planes of the complex variable $x$. The functions $f_{\pm }$ can be written as projectors
$\hat{P}_{\pm }=(1\mp i\hat{H})/2$ actions on the function $f$: $f_{\pm 
\text{ }}=\hat{P}_{\pm }f$. Due to the projectors properties, $\hat{P}_{\pm }^{2}=\hat{P}_{\pm }$ and $\hat{P}_{\pm }\hat{P}_{\mp }=0$, which are equivalent to the condition $\hat{H}^{2}=-1$, the nonlinear terms on the right-hand side of Eq.~\eqref{eqkey1} split into a sum of functions  analytically continuable into the upper and lower half-planes: 
\begin{eqnarray*}
(\hat{k}f)^{2}-(f_{x})^{2} &\equiv &(\hat{H}f_{x})^{2}-(f_{x})^{2}=-2\left(
f_{+x}^{2}+f_{-x}^{2}\right) , \\
f_{x}\hat{k}f &\equiv &-f_{x}\hat{H}f_{x}=-i\left(
f_{+x}^{2}-f_{-x}^{2}\right) .
\end{eqnarray*}
It is clear that the linear terms in Eq.~\eqref{eqkey1} are also can be represented as a sum of functions analytically  continuable into the upper and lower half-planes. Therefore, the equations for $f_{\pm }$ are separated into two independent equations. In particular, for $f_{+}\equiv F$, we have an autonomous equation which does not contain $f_{-}$: 
\begin{equation}
F_{t}+iF_{x}=-e^{i\gamma }F_{x}^{2},  
\label{eqkey2}
\end{equation} 
where  we have introduced notation $\gamma =\mbox{arccos}{A}$. The real parameter $\gamma$, defined by the Atwood number, lies in the range $0\leq \gamma \leq \pi $. So, the values $A=1,\,0,\,-1$ correspond to $\gamma
=0,\,\pi /2,\,\pi $. The similar equation could be obtained for $f_{-}$ (it coincides with the complex conjugate of Eq.~\eqref{eqkey2}).

It is important that, unlike the original equations, the equation \eqref{eqkey2} is local, i.e., it does not contain integral operators. 
Differentiating this equation with respect to $x$ leads to the equation of the Hopf-type:  
\begin{equation}
V_{t}+iV_{x}=-2e^{i\gamma }VV_{x},  \label{eqkey3}
\end{equation}
where $V=F_{x}$ has a meaning of  the complex velocity. The solution to this equation can be found by means of the method of characteristics,  
\begin{equation}
V=V_{0}(\tilde{x}),\quad \quad x=\tilde{x}+it+2e^{i\gamma }V_{0}(\tilde{x})t,
\label{eqreshf}
\end{equation}
where the function $V_{0}$ is defined from the initial condition $V_{0}(x)=V|_{t=0}$, and $\tilde{x}$ has a meaning of the Lagrangian coordinate. Since the function $V$ is analytic in the upper half-plane, all its singularities are located in the lower half-plane. As was shown in Refs.~\cite{kuz1, kuz2}, 
every point singularity transforms at $t>0$ into a cut with the movable branch points. Their locations are defined by the condition $\partial x/\partial \tilde{x}=0$, i.e., 
\begin{equation}
1+2e^{i\gamma }V_{0}^{\prime }(\tilde{x})t=0  \label{eqbranch1}
\end{equation}
(here the subscript denotes the derivative with respect to the argument). At the moment of time $t=t_{c}$, when the most rapid branch point reaches the real axis, the analyticity of $V$, obviously, breaks down and, hence, a singularity appears in the solution. Let us consider this situation in more detail.

The equation \eqref{eqbranch1} defines the trajectories along which the branch points move over the complex $\tilde{x}$ plane. Let $\tilde{x}=\tilde{X}(t)$ be one of these trajectories. According to \eqref{eqreshf}, the motion of the corresponding branch point in the complex $x$ plane is described by the equation 
\[
x=X(t)=\tilde{X}(t)+it+2e^{i\gamma }V_{0}\left( \tilde{X}(t)\right) t. 
\] 
Let this point reach the real axis first; the moment of singularity formation is then defined from the equality 
$\mbox{Im}\,X(t_{c})=0$. The expansion of the expressions \eqref{eqreshf} in a neighborhood of the point $t=t_{c}$, $x=x_{c}\equiv X(t_{c})$, 
$\tilde{x}=\tilde{x}_{c}\equiv \tilde{X}(t_{c})$ gives in the leading order 
\[
V=V_{0}(\tilde{x}_{c})+V_{0}^{\prime }(\tilde{x}_{c})\delta \tilde{x}+\ldots, \]
\[
\delta x=i\delta t+2e^{i\gamma }V_{0}(\tilde{x}_{c})\delta t+t_{c}e^{i\gamma
}V_{0}^{\prime \prime }(\tilde{x}_{c})(\delta \tilde{x})^{2}+\ldots , 
\]
where $\delta t=t-t_{c}$, $\delta x=x-x_{c}$, and $\delta 
\tilde{x}=\tilde{x}-\tilde{x}_{c}$. Excluding the parameter $\delta \tilde{x}$, we obtain 
\begin{equation}
V(x,t)=V_{0}(\tilde{x}_{c})+V_{0}^{\prime }(\tilde{x}_{c})\left[ 
\frac{\delta x-\left( i+2e^{i\gamma }V_{0}(\tilde{x}_{c})\right) \delta 
t}{t_{c}e^{i\gamma }V_{0}^{\prime \prime }(\tilde{x}_{c})}\right] ^{1/2}+\ldots
\label{eqrazl1}
\end{equation}
Hence one can see that the derivatives $V_{x}$ and $V_{t}$ become singular. As a result, the boundary shape acquires root singularities. In particular,   from \eqref{eqrazl1} follows 
\begin{equation}
V_{x}(x,t)\approx \frac{V_{0}^{\prime 
}(\tilde{x}_{c})}{2\sqrt{t_{c}e^{i\gamma }V_{0}^{\prime \prime }(\tilde{x}_{c})}}\left[ \delta
x-\left( i+2e^{i\gamma }V_{0}(\tilde{x}_{c})\right) \delta t\right] ^{-1/2},
\label{eqrazlx1}
\end{equation}
i.e., in the general case,  
$V_{x}(x_{c},t)\sim \,|\delta t|^{-1/2}$.

Thus, we have reduced the nonlocal equation \eqref{eqkey1} for the function $f$ to the local partial differential equation of the first order \eqref{eqkey2} for the function $F$ analytical in the upper half-plane of complex variable $x$. 
It turned out that this equation admits exact analytical solution. 
In order to find the interface shape, which is determined as $\eta =f-g$, it is necessary to know not only the function $f$ but also the function $g$, which can be found by solving Eq.~\eqref{eqrel1}. In the linear approximation, this equation describes the relaxation of $g$ to zero, however, the presence of the quadratic in $f$ terms in the right-hand side of Eq.~\eqref{eqrel1} leads to the induced growth of the function $g$. 
For small angles of the surface inclination, the value of $g$ will be of the order of $O(f^{2})$ and, therefore, its role will be insignificant for  weakly nonlinear evolution of this system.  
However, this statement requires to be verified in the vicinity of the singular point where the second derivative of the function $f$ becomes singular. 

Introducing the analytical continuation of the function $g$ into the upper half-plane, $G=\hat{P}_+g$, we rewrite Eq.~\eqref{eqrel1} as 
\begin{equation}
G_{t}-iG_{x}=-e^{i\gamma }F_{x}^{2}+2e^{-i\gamma }\hat{P}_+(F\bar{F}_{x})_{x},
\label{eqrel2}
\end{equation}
where $\bar{F}$ means the complex conjugate of $F$. 
The function $\bar{F}$ is analytical in the lower half-plane; its singularities lie in upper $x$ half-plane. If Eq.~\eqref{eqrel2} does not contain the projection operator $\hat{P}_+$, the second term in the right-hand side would be singular due to the second derivative $\bar{F}$; however the action of $\hat{P}_+$ suppresses the appearance of a singularity at $\mbox{Im}\,x>0$. 
Hence it is easy to understand that, due to the projector, the term with the second derivative of $\bar{F}$ will everywhere have the same order as the first term in the right-hand side of Eq.~\eqref{eqrel2}, including the neighborhood of the touching point $x=x_{c}$. 
In the neighborhood of this point, the function $G$ has to follow the asymptotics (\ref{eqrazl1}), i.e., $G$ should be sought in the form 
\[
G=G\left(\delta x-\left( i+2e^{i\gamma }V_{0}(\tilde{x}_{c})\right) \delta
t\right). 
\]
In this case, we obtain from Eq.~(\ref{eqrel2}) 
\[
-2\left[i+e^{i\gamma}V_{0}(\tilde{x}_{c})\right]G_{x} =-e^{i\gamma
}F_{x}^{2}+2e^{-i\gamma }\hat{P}_+(F\bar{F}_{x})_{x}. 
\]
The real part of $G_{x}$ gives the estimate for the contribution to the characteristic slope angle. 
It is clear that it is small compared to the real part of $F_{x}$, i.e., the surface shape near the singularity is determined by the function $F$ in the small-angle approximation: $\eta\simeq 2\mbox{Re}\,F$. 
This function defines the corresponding singularities of the surface $y=\eta(x,t)$ for $t\to t_c$.

Let us now discuss the interface dynamics. In the framework of the quadratic-nonlinear approximation, the interfacial curvature, specified as $\eta _{xx}(1+\eta _{x}^{2})^{-3/2}$, is of
\begin{equation}  \label{eqcurfg}
\eta _{xx} \approx 2\,\mbox{Re}\,F_{xx}=2\,\mbox{Re}\,V_{x}.
\end{equation}
We will assume that the condition $|V_{0} (\tilde{x}_{c} )|\ll 1$ holds. It is realized in the case where the small-angle approximation is valid (the function $V_{0}(\tilde{x}_{c})$ defines the interface inclination  at the singular point). Then the expansion \eqref{eqrazl1} for the function $V$ in the neighborhood of the singular point $x=x_{c}$ and $t=t_{c}$ can be approximately written as 
\begin{equation}  \label{eqrazlun}
V(x,t)\approx V_{0} (\tilde{x}_{c} )+V^{\prime }_{0} (\tilde{x}_{c} 
)\left[\frac{\delta x-i\delta t}{t_{c} e^{i\gamma } V^{\prime \prime }_{0} 
(\tilde{x}_{c} )} \right]^{1/2} .
\end{equation}
It can be seen that, close to the singularity, the behavior of the system becomes universal: the function $V$ depends only on the combination of variables $(\delta x-i\delta t)$ for arbitrary initial conditions. The initial conditions determine only the additive constant and the constant factor in the expression \eqref{eqrazlun}. Such a dependence is characteristic for solutions of the linearized equation \eqref{eqkey3}: 
\begin{equation}  \label{eqlinear}
V_{t} +iV_{x} =0.
\end{equation}
Substituting \eqref{eqrazlun} into this equation yields  identity. 
Of course, it does not mean that the singularity formation can be described within the linear equation \eqref{eqlinear}. The nonlinear term of Eq.~\eqref{eqkey3} defines the type of a forming singularity, specifying a concrete dependence for  $V$ as a function of the combination $(\delta x-i\delta t)$. In particular, just the nonlinearity  determines  influence of the Atwood number on the system behavior: the linearized equation \eqref{eqlinear} does not contain the $A$ number (or, which is the same, the parameter $\gamma$) in the explicit form.

We return to the discussion of the boundary behavior close to the singular point. 
Using \eqref{eqrazlun}, we find the following universal relationship for the curvature in the singular point vicinity: 
\[
\eta _{xx}\approx \,\mbox{Re}\left\{ \frac{V_{0}^{\prime 
}(\tilde{x}_{c})}{\sqrt{t_{c}e^{i\gamma }V_{0}^{\prime \prime }(\tilde{x}_{c})\left( \delta
x-i\delta t\right) }}\right\} .
\]
Note that some of our conclusions are based on the small-angle approximation, which validity at the singular moment  $t=t_{c}$ requires additional verification.

\section{Evolution of periodic perturbations of the interface}

Let at the initial moment $t=0$ 
\[
F(x,0)=-ia_{0}e^{ix},\quad \quad G(x,0)=0,
\]
that corresponds to the periodic perturbation of the interface (with  period $2\pi $) 
\[
\eta (x,0)=2a_{0}\sin x.
\]
Here $2a_{0}$ is the initial amplitude of the interface deformation which is assumed small, $0<a_{0}\ll 1$. These initial conditions correspond to  
\begin{equation}
V(x,0)=a_{0}e^{ix}.  \label{eqinit}
\end{equation} 
The evolution of the function $V$, according to the solution 
\eqref{eqreshf}, is determined by the expressions 
\begin{equation}
V=V_{0}(\tilde{x})=a_{0}e^{i\tilde{x}},\quad \quad 
x=\tilde{x}+it+2a_{0}te^{i\gamma +i\tilde{x}}.  \label{eqreshper}
\end{equation}
Equation \eqref{eqbranch1}, which defines locations for  the $\tilde{x}(x)$ mapping singularities, takes the following form: 
\begin{equation}
1+2ia_{0}e^{i\gamma +i\tilde{x}}t=0.  \label{eqbranch2}
\end{equation}
In accordance with it, the singularities move in the  $\tilde{x}$ plane along the trajectories 
\[
\tilde{x}=\tilde{X}_{n}(t)=\pi /2+2\pi n-\gamma +i\ln (2a_{0}t),
\] 
where $n$ is an integer number. Substituting this expression into \eqref{eqreshper}, we find that the branch points of the function $V$ move over the lower half-plane of the complex variable $x$ towards the real axis along the straight lines parallel to the imaginary axis, 
\[
x=X_{n}(t)=\pi /2+2\pi n-\gamma +i\ln (2a_{0}t)+it+i.
\]
Singularities on the interface appear (simultaneously) at the moment $t=t_{c}$ when the branch points reach the imaginary axis: $\mbox{Im}\,X_{n}(t_{c})=0$. Restricting by a single spatial period $-\pi \leq x\leq \pi $ ($n=0$), we find 
\begin{equation}
x_{c}=\pi /2-\gamma ,\qquad \ln (2a_{0}t_{c})+t_{c}+1=0.  \label{eqsing}
\end{equation}
One can see that the coordinates of the singular points depend on the parameter $\gamma $ and, therefore, on the Atwood number. 
So, we have $x_{c}=\pi /2$ for $\gamma =0$ ($A=1$), i.e., the singularities appear at the maximums of the function $\eta$; we get $x_{c}=-\pi /2$ for $\gamma =\pi $ ($A=-1$), i.e., the singularities arise at the minimums of $\eta$. For intermediate values of $A$, the singularities lie in the range $-\pi /2<x_{c}<\pi /2$. In particular, $x_{c}=0$ for $\gamma =\pi /2$ ($A=0$) that corresponds to the inflection point of the function $\eta$. The time $t_{c}$ does not depend on the parameter $A$: it depends only on the initial amplitude $a_{0}$. Since $a_{0}\ll 1$, the solution of the transcendental equation \eqref{eqsing} for $t_{c}$ can be found by iterations:  
\begin{equation}
t_{c}=-\ln a_{0}-\ln (-\ln a_{0})-\ln 2-1+\ldots .  \label{eqtime1}
\end{equation} 
Thus, $t_{c}\gg 1$ for sufficiently small amplitude $a_{0}$. This means that the most of  time, until $t=t_{c}$, the system is in the stage of linear instability. Nonlinear effects occur as $t$ approaches $t_{c}$. It should be noted that at $A=0$  equations \eqref{eqsing} coincide with the results concerning the instability of a vortex sheet \cite{moore,saf}.

In the previous section, for demonstrating the behavior universality near the singular point, we have used the condition $|V_{0}(\tilde{x}_{c})|\ll 1$. 
It can be seen from Eqs.~\eqref{eqinit} and \eqref{eqbranch2} that $V_{0}(\tilde{x}_{c})=ie^{-i\gamma }/2t_{c},$ namely, the parameter $V_{0}(\tilde{x}_{c})$ is small in absolute value for sufficiently large  $t_{c}$ that, according to \eqref{eqtime1}, corresponds to 
$a_{0}\rightarrow 0$. In particular, this implies smallness of the interface slope angle  at the singular point; it equals approximately
$2\,\mbox{Re}V_{0}(\tilde{x}_{c})=t_{c}^{-1}\sin \gamma$.

Let us show that, at the moment of singularity formation, the slope angles of the interface are small not only in the neighborhood of the singularity but also at the periphery. Accounting for \eqref{eqsing}, the expressions \eqref{eqreshper} for the function $V$ at the moment $t=t_{c}$ can be presented as 
\begin{equation}
V=a_{0}e^{t_{c}+i\xi },\quad \quad x=\xi +e^{i\gamma +i\xi -1},
\label{eqreshperc}
\end{equation} 
where we have introduced the auxiliary variable $\xi \equiv \tilde{x}+it_{c}$. 
It can be seen that the second equation, determining the mapping $\xi(x)$, does not contain the small parameter $a_{0}$, and also the related parameter $t_{c}$. Hence it follows immediately  that $\xi =O(1)$ for $-\pi \leq x\leq \pi $ (there is no need to solve this transcendental equation). 
One can see from the first equation of \eqref{eqreshperc} that the characteristic boundaryslope angles, $\eta _{x}\approx 2\,\mbox{Re}V$, are defined by the factor $a_{0}e^{t_{c}}$. Taking into account \eqref{eqsing}, we get $a_{0}e^{t_{c}}=(2et_{c})^{-1}$. Then, as follows from \eqref{eqtime1}, the slope angles of the interface are small at 
$a_{0}\rightarrow 0$.

Thus, the analysis of the evolution for  periodic perturbations with a small amplitude, presented in this section, has demonstrated that the small-angle approximation does not break down at the moment of singularity formation, that proves the applicability of our approach based on expansions with respect to the canonic functions $\psi $ and $\eta $, or with respect to the auxiliary functions $f$ and $g=O(f^{2})$.

Let us now discuss the type of singularities appearing on the interface. 
It follows from \eqref{eqrazlun} that  
\begin{equation}
V_{x}(x,t)\approx -\frac{e^{-i\gamma }}{2t_{c}}\left[ -2\delta t-2i\delta 
x\right] ^{-1/2}  \label{eqcurv00}
\end{equation}
in the neighborhood of the singular point. Here we have used the relations
$V_{0}(\tilde{x}_{c})=(i/2t_{c})e^{-i\gamma }$ and  
$V_{0}^{\prime \prime }(\tilde{x}_{c})=-(i/2t_{c})e^{-i\gamma }$. 
It can be seen that the derivative $V_{x}$ tends to infinity according to the law $V_{x}(x_{c},t)\sim (-\delta t)^{-1/2}$. As discussed above, it is just the function $V$ that defines the interface behavior, and, therefore, the curvature in the leading order behaves as $\eta _{xx}\approx 2\,\mbox{Re}V_{x}$ . Separating the real part of \eqref{eqcurv00}, we get  
\begin{equation}
\eta _{xx}\approx -\frac{A\left( \sqrt{\delta t^{2}+\delta x^{2}}-\delta
t\right) ^{1/2}+\mbox{sgn}(\delta x)\sqrt{1-A^{2}}\left( \sqrt{\delta
t^{2}+\delta x^{2}}+\delta t\right) ^{1/2}}{2t_{c}\sqrt{\delta t^{2}+\delta
x^{2}}},  \label{eqcurvnew}
\end{equation} 
where we have returned to using the parameter $A$. 
This expression takes the most simple form for $A=\pm 1$ and for 
$A=0$, when one of two terms in the numerator vanishes. 
So, at $A=\pm 1$, the curvature $\eta _{xx}$ becomes an even function of  $\delta x$. At the singular point $x_{c}=\pm \pi /2$, the curvature becomes infinite in a finite time: 
$\eta _{xx}(x_{c},t)\approx \mp t_{c}^{-1}(-2\delta
t)^{-1/2}$.
It is negative for $A=1$ and positive for $A=-1$.

For $A=0$ the curvature $\eta _{xx}$ becomes an odd function of  
$\delta x$. The curvature is equal to zero at the singular point $x_{c}=0$, i.e., we deal with the inflection point. At the  moment of singularity formation $\eta _{xx}(\pm 0,t_{c})\rightarrow \mp \infty $, namely, the curvature has a discontinuity of the second kind. It should be noted that such a behavior of the interface was predicted by Moore on the basis of  the analysis of the evolution for periodic perturbations of the vortex sheet \cite{moore,saf}.

At arbitrary $A$ 
\[
\eta _{xx}(x,t_{c})\approx -\frac{A+\mbox{sgn}(\delta 
x)\sqrt{1-A^{2}}}{2t_{c}\sqrt{|\delta x|}}
\]
at the moment $t=t_{c}$ in the vicinity of the singular point. 
It can be seen that the curvature sign  is defined by the expression $\left( -A-\sqrt{1-A^{2}}\right)$ to the right of the singular point, and by the expression $\left( -A+\sqrt{1-A^{2}}\right)$ to the left of it.  
Hence, if the Atwood number varies from the minimum possible value (-1) up to the maximum one (+1), the curvature changes its sign from positive to negative at $A=-1/\sqrt{2}$ to the right of the singular point, and at $A=1/\sqrt{2}$ to the left of this point. 
Then we can conclude that the curvature is always negative close to the singular point for $1/\sqrt{2}<A\leq 1$ (i.e., for 
$\rho _{1}/\rho _{2}>(\sqrt{2}+1)/(\sqrt{2}-1)\approx 5.83$), and it is always positive for $-1\leq A<-1/\sqrt{2}$ (i.e., for 
$\rho _{2}/\rho _{1}>5.83$). 
In the intermediate case of $-1/\sqrt{2}<A<1/\sqrt{2}$, the curvature changes the sign at the singular point.

\section{Evolution of the interface localized perturbations }

Now consider the interface dynamics  for spatially localized perturbations, i.e., $\eta \rightarrow 0$ and $\psi \rightarrow 0$ at $|x|\rightarrow \infty $.

Let
\begin{equation}
V(x,0)=-\frac{ise^{i(\alpha -\gamma )}}{(x+i)^{2}}  \label{eqinitloc}
\end{equation}
at $t=0$, namely, at the initial moment of time the function $V$ has a pole of the second order at the point $x=-i$. Here the condition of the small-angle approximation, $s\ll 1$, is assumed to be satisfied for the constant $s$; $\alpha $ is a real parameter lying in the range $-\pi \leq \alpha \leq \pi $. We also suppose that $G(x,0)=0$; then the shape of the interface has the following form at $t=0$:
\[
\eta (x,0)=\frac{2s\cos (\alpha -\gamma )}{x^{2}+1}-\frac{2sx\sin (\alpha
-\gamma )}{x^{2}+1}.
\]

In accordance with \eqref{eqreshf}, the dynamics of the function $V$ is defined by the expressions
\begin{equation}
V=V_{0}(\tilde{x})=-\frac{ise^{i(\alpha -\gamma )}}{(\tilde{x}+i)^{2}},
\label{eqreshloc1}
\end{equation}
where  
\begin{equation}
x=\tilde{x}+it-\frac{2iste^{i\alpha },}{(\tilde{x}+i)^{2}}.
\label{eqreshloc2}
\end{equation}
The equation \eqref{eqbranch1}, describing the motion of branch points in the  $\tilde{x}$ plane along the trajectories $\tilde{x}=\tilde{X}(t)$, 
takes the form 
\[
1+\frac{4iste^{i\alpha }}{(\tilde{X}(t)+i)^{3}}=0.
\]
This equation has three roots, 
\begin{equation}
\tilde{X}_{n}(t)=-i+ie^{i(\alpha +2\pi n)/3}(4st)^{1/3},\quad \quad
n=1,\;2,\;3.  \label{eqbrtil123}
\end{equation} 
Then the trajectories of the branch points in the $x$ plane, resulting from \eqref{eqreshloc2}, are given by the expressions
\begin{equation}
x=X_{n}(t)=it-i+3ie^{i(\alpha +2\pi n)/3}(st/2)^{1/3},\qquad n=1,\;2,\;3.
\label{eqbranch123}
\end{equation}
The point $X_{3}$ first reaches the real axis: the nonlinearity accelerates its motion (linear solutions correspond to the limit $s\rightarrow 0$). Note that, in the particular case of $\alpha =\pi$, the point $X_{3}$ touches the real axis simultaneously with the point $X_{2}$, and, in the case of $\alpha =-\pi $, it touches the real axis simultaneously with the point $X_{1}$.

The moment of the singularity formation is determined by the condition $\mbox{Im}\,X_{3}(t_{c})=0$: 
\[
t_{c}-1+3\cos (\alpha /3)(st_{c}/2)^{1/3}=0.
\] 
Since the value of $s$ is small, we have for $t_{c}$ approximately: 
\begin{equation}
t_{c}\approx 1-3\cos (\alpha /3)(s/2)^{1/3}.  \label{eqtimec}
\end{equation} 
According to \eqref{eqbranch123}, the branch point $x=X_{3}(t)$ reaches the real axis at the point 
\[
x=\mbox{Re}\,X_{3}(t_{c})\approx -3\sin (\alpha /3)(st_{c}/2)^{1/3}.
\] 
In the particular case when $\alpha =0$ the branch point moves along the imaginary axis, and the singularity appears at the point $x=0$.

Let us consider the type of the forming singularity. In Sec.~5 we have shown that, in the general case, the behavior of the system near the singularity is defined by the parameters $V_{0}(\tilde{x}_{c})$, $V_{0}^{\prime 
}(\tilde{x}_{c})$, and $V_{0}^{\prime \prime }(\tilde{x}_{c})$ which, in turn, are determined by the initial conditions. Substituting the expression for time \eqref{eqtimec} into \eqref{eqbrtil123} we find for $\tilde{x}_{c}$: 
\[
\tilde{x}_{c}\equiv \tilde{X}_{3}(t_{c})=-i+ie^{i\alpha
/3}(4st_{c})^{1/3}\approx -i+ie^{i\alpha /3}(4s)^{1/3}.
\]
Further, we obtain the relations for the desired parameters from \eqref{eqreshloc1}: 
\[
V_{0}(\tilde{x}_{c})=ie^{i(\alpha /3-\gamma )}s^{1/3}/2^{4/3},\qquad
V_{0}^{\prime }(\tilde{x}_{c})=-e^{-i\gamma }/2,\qquad V_{0}^{\prime \prime
}(\tilde{x}_{c})=-3ie^{-i(\alpha /3+\gamma )}s^{-1/3}/2^{5/3}.
\]
Note that $|V_{0}(\tilde{x}_{c})|\sim s^{1/3}$ and, as a consequence, $|V_{0}(\tilde{x}_{c})|\ll 1$. This condition was used earlier by us for deriving the expansion \eqref{eqrazlun}. As a result, we get from \eqref{eqrazlx1} the expression 
\[
V_{x}(x,t)\approx -\frac{s^{1/6}e^{+i(\alpha /6-\gamma 
)}}{3^{1/2}2^{5/6}}\left[ -\delta t-i\delta x\right] ^{-1/2},
\] 
that differs only by a constant factor from \eqref{eqcurv00}. 
Hence we obtain for the interface curvature, 
\begin{equation}
\eta _{xx}\approx -\frac{\cos (\gamma -\alpha /6)\left( \sqrt{\delta
t^{2}+\delta x^{2}}-\delta t\right) ^{1/2}+\mbox{sgn}(\delta x)\sin (\gamma
-\alpha /6)\left( \sqrt{\delta t^{2}+\delta x^{2}}+\delta t\right) 
^{1/2}}{3^{1/2}2^{5/6}s^{-1/6}\sqrt{\delta t^{2}+\delta x^{2}}}.  \label{eqcurv}
\end{equation}
This means that the instability development results in the formation of the singular point in which the curvature becomes infinite in a finite time. At the moment $t_{c}$, we have in the neighborhood of the point $x_{c}$: 
\[
\eta _{xx}(x,t_{c})\approx -\frac{\cos (\gamma -\alpha /6)+\mbox{sgn}(\delta
x)\sin (\gamma -\alpha /6)}{3^{1/2}2^{5/6}s^{-1/6}\sqrt{|\delta x|}}.
\]
It follows from this expression that, in particular, the curvature will be negative near the singular point for any $\alpha$ for 
$0\leq \gamma <\pi /12$ (this corresponds to the Atwood number in the range $\cos (\pi /12)<A\leq 1$ or, what is the same, 
$\rho_{1}/\rho _{2}>
(4+\sqrt{6}+\sqrt{2})/(4-\sqrt{6}-\sqrt{2})\approx 57.7$)
and positive for any $\alpha$ for $11\pi /12\leq \gamma <\pi $ 
(this corresponds to $-1\leq A<-\cos (\pi /12)$ and 
$\rho _{2}/\rho _{1}>57.7$). 
In other cases, the curvature can change its sign in the singular point. The curvature changes the sign for any $\alpha$ for $5\pi /12<\gamma <7\pi /12$ (this corresponds to the Atwood number in the range $\cos (7\pi /12)<A<\cos (5\pi /12)$ or, what is the same, 
$0.59\approx 
(4-\sqrt{6}+\sqrt{2})/(4+\sqrt{6}-\sqrt{2})<
\rho _{1}/\rho _{2}<(4+\sqrt{6}-\sqrt{2})/(4-\sqrt{6}+\sqrt{2})
\approx 1.70$). 
These results are consistent with those from Sec.~6, where the periodic interface perturbations are considered.

Since $\mbox{Re}V_{0}(\tilde{x}_{c})\sim s^{1/3}$, the slope interface angle  is small in the singular point. Let us show that the characteristic slope angles at the moment $t=t_{c}$ will be small also in the periphery, i.e., outside the vicinity of the singular point $x=x_{c}$. Recall that the condition of smallness of slope angles was used when we derived the key equation \eqref{eqkey3}.

Excluding some spatial-temporal neighborhood of the singularity, where everything is determined by the influence of nonlinearity, the system dynamics can be approximated  by the linearized equation \eqref{eqlinear}. Its solution with the initial condition \eqref{eqinitloc} is 
\begin{equation}
V(x,t)=-\frac{ise^{i(\alpha -\gamma )}}{\left[ x+i(1-t))\right] ^{2}},
\label{eqreshlinloc}
\end{equation}
that is the pole moves along the imaginary axis to the origin with constant velocity. At the moment $t=1$, a strong singularity develops at the point $x=0$, violating the small-angle approximation. However, as seen from \eqref{eqtimec}, the influence of nonlinear terms accelerates the formation of singularity and, consequently, the angles will remain finite at the moment of its formation. Indeed, substituting the formula \eqref{eqtimec} for the time $t_{c}$ into \eqref{eqreshlinloc}, we obtain  
\[
V(x,t_{c})\approx -\frac{ise^{i(\alpha -\gamma )}}{\left[ x+3i\cos (\alpha
/3)\;(s/2)^{1/3}\right] ^{2}}.
\]
Hence it follows that the characteristic value of the derivative $\eta _{x}$ will be of the order of $s^{1/3}$, i.e., the angles remain small, and the applicability of our expansions is not violated over the entire time interval 
$0\leq t\leq t_{c}$.

\section{Conclusion}

In the present work it was demonstrated that, due to the Kelvin-Helmholtz instability, the evolution of the interface between two fluids moving relative to each other results in appearance of weak (root) singularities for the interfacial curvature of two types  \eqref{eqcurvnew} and \eqref{eqcurv} in a finite time. 
For periodic perturbations of the interface with initially small slope angles of the order of $a_{0}$, the angles reach values of the order  $|1/\ln {a_{0}}|$, i.e., they remain small at the moment of singularity formation. 
For localized perturbations the slope angles of the surface  $y=\eta (x,t)$ increase from the initial values of the order of $s$ up to the values of  order  $\sim s^{1/3}$ at $t=t_{c}$ still remaining small. 
Thus, the condition for small slope angles, which was used for deriving the key equations, is not violated.

At the moment of singularity formation, $t_{c}$, the interfacial curvature in the vicinity of the singular point  $x_{c}$ is defined by  
\[
\eta _{xx}(x,t_{c})\approx |x-x_{c}|^{-1/2}\left( 
c_{1}+c_{2}\mbox{sgn}(x-x_{c})\right) ,
\]
where the constants $c_{1,2}$ are defined by the initial perturbation shape of  the interface and also by Atwood number $A$. If $|c_{1}|<|c_{2}|$, then the curvature changes its sign at the singular point. 
Otherwise, it has a definite sign. 
Our analysis has shown that, in the general case, for  fluids with comparable densities (Atwood number is small in absolute value, $|A|\ll 1$), the curvature always changes its sign at the singular point. 
The classic situation when the fluids are identical ($A=0$) corresponds to this case. 
If the density of one fluid considerably exceeds the density of another fluid (i.e., for Atwood number close to unity, $|A|\approx 1$), the interfacial curvature has a definite sign near the singularity: it is negative for $A\approx 1$ and positive for $A\approx -1$.

It is noteworthy that, in the limits $A\to\pm 1$, the evolution of the Kelvin-Helmholtz instability is similar to the behavior of the free surface of a conducting liquid in a strong vertical electric field. 
As was demonstrated in Refs.~\cite{zub1,zub2}, there is a tendency for the formation of the sign-definite singularities of the type of 
$\eta_{xx}\sim |x-x_c|^{-1/2}$ on the boundary of a conducting liquid at the nonlinear stage of the Tonks-Frenkel instability.  
Note also that similar properties are demonstrated by the interface between two ideal dielectric fluids in a strong vertical electric field \cite{zub3}.

\medskip 

This study was supported by the program ``Fundamental Problems of Nonlinear Dynamics in Mathematical and Physical Sciences'' of the Presidium of RAS. 
The work of N.M.Z. was partially supported by the Russian Foundation for Basic Research and by the Government of Sverdlovsk Region (grant no. 13-08-96010-Ural). 
The work of E.A.K. was supported by a grant from the government of
Russia for the state support of scientific research, conducted
under the supervision of leading scientists in Russian
educational institutions in the system of high professional
education (Contract 11.G34.31.0035 from 25 November 2010
between the Ministry for Education and Science, Novosibirsk
State University, and the leading scientist), as well as by the Russian Foundation for Basic Research (grant no. 12-01-00943) and a grant from leading scientific schools of
Russia, NSh 3753.2014.2.

\end{document}